\documentclass{article}

\usepackage[utf8]{inputenc}
\usepackage{amssymb}
\usepackage{url}
\usepackage{nopageno}
\usepackage{xcolor}
\usepackage{graphicx}
\usepackage{enumerate}
\usepackage{array}
\usepackage{ragged2e}
\usepackage{amsmath}
\usepackage{appendix}
\usepackage{authblk}
\usepackage[style=oscola]{biblatex}
\usepackage{lineno}

\begin{document}

\title{Dissecting liabilities in adversarial surgical robot failures: A national (Danish) and EU law perspective}

\author[1]{K. Rosager Ludvigsen}
\author[2]{Shishir Nagaraja}
\affil[1]{Department of Computer and Information Sciences, University of Strathclyde, kaspar.rosager-ludvigsen@strath.ac.uk}
\affil[2]{Department of Computer and Information Sciences, University of Strathclyde, shishir.nagaraja@strath.ac.uk}
\date{January 2022}

\maketitle

\begin{abstract}

Over the last decade, surgical robots have risen in prominen2ce and usage. They are not merely tools, but have also become advanced instruments with network connectivity. Connectivity is necessary to accept software updates, accept instructions, and transfer sensory data, but it also exposes the robot to cyberattacks, which can damage the patient or the surgeon. These injuries are normally caused by safety failures, as seen in accidents with industrial robots, but cyberattacks are caused by security failures instead. We create a taxonomy for both types of failures in this paper specifically for surgical robots. These robots are increasingly sold and used in the European Union (EU), hence it is natural to consider how surgical robots are viewed and treated by EU law. Specifically, which rights regulators and manufacturers have under it, and which legal remedies and actions a patient or manufacturer would have in a single national legal system in the union, if injuries were to occur from a security failure caused by an adversary that cannot be unambiguously identified (attribution of cyberattacks is often hard). Given that the Medical Device Regulation (MDR) has only recently entered into force, we also offer some general considerations of the regulation.

We find that the selected (Danish) national legal system can adequately deal with attacks on surgical robots, because it can on one hand efficiently compensate the patient, and at the same time protect the patient by not shying away from dealing with the problem concretely. This is because of its flexibility; secondly, a remarkable absence of distinction between safety vs security causes of failure and focusing instead on the detrimental effects, thus benefiting the patient; and third, liability can be removed from the manufacturer by withdrawing its status as party, if the patient chooses a separate public law measure to recover damages. Furthermore, we find that current EU law does consider both security and safety aspects of surgical robots, without it mentioning it through literal wording, but it also adds substantial liabilities and responsibilities to the manufacturers of surgical robots, gives the patient special rights and confers immense powers on the regulators, which can end up affecting any future lawsuits.
\end{abstract}

\section{Introduction}

In a world with an increased use of robot technology, surgeons closely work with robots in a myriad of roles, from offering assistance to leading surgical procedures. Surgical robots enable unique approaches to treatment not possible before, with minimally invasive surgery being the primary technique (robotic surgeons can work more efficiently than human surgeons in smaller spaces due to highly accurate actuation). Two examples would be laparascopy done by the da Vinci systems\footnote{Operation by small incisions into abdomen or pelvis with the aid of a camera.} and the Magellan system deployed for cardiac surgery.\footnote{\cite[3]{Bergeles2014}} Surgical robots are widely used for a range of treatments, \footcite[388]{Holder2016} for example for hernia and intestinal cancer.\footnote{Eg a da Vinci surgical robot from a small Danish hospital broke the record with 426 performed surgeries in 2019, and it is mainly used for the aforementioned treatments, see \url{<https://jv.dk/artikel/vild-statistik-på-sygehus-robot-står-bag-1000-operationer>}, last accessed 14 December 2021.} Their increased usage are not without consequences, and criticisms of specific surgeries has started to occur, primarily aimed at the lack of empirical research that makes their efficiency likely.\footnote{See eg \cite{Dhanani2021}.} We see this development increased around the world, especially in the US and the EU.\footnote{eg \url{<https://www.medtechdive.com/news/intuitive-surgical-profit-up-on-strong-da-vinci-robot-sales/528257/>}, last accessed 14 December 2021.}

The surgeon and the patient are not the only relevant parties when it comes to surgical robots. Engineers, programmers, nurses, lawyers and a whole range of other staff are needed to design, produce, operate and maintain surgical robots,\footnote{This is especially important if they end up being partially or fully controlled by AI, which may have a negative social impact, see eg \cite[5]{Gomez-Gonzalez2020}.} and must handle and possibly mitigate accidents if they happen. 
\newpage

Certain research on the perspective of manufactures and patients has been done in the area,\footnote{See the note by \cite{Beglinger2020}, for an example of an interdisciplinary approach with focus on liability in a US context.} but it is still underdeveloped regarding the legal rights, liabilities, obligations and the choice of legal instruments and tools that are applied to surgical robots in general. It is this space to which we would like to contribute.

For us to legally discuss surgical robots, we must categorise them and the features of which they consist of. We choose to consider them as cyberphysical systems (CPS),\footcite[2]{Solicitation2014} because they are robots that interface with the physical world,\footnote{There is no doubt that many types of medication prescription systems, or those that monitor the health of the patient (but do nothing else), will be considered CPS even if they barely interface with the physical world.} with their tools being used directly on the patient. Robots can be defined in variety of ways, but we choose to focus on its CPS nature.\footnote{Other ways could be those proposed in \cite[13]{Fosch-Villaronga2019} which apply to robots in general and could be appropriate.}

CPS has a tendency to erase the boundary between the physical and digital sphere, which is clearly seen in the many new ways which they can injure people or otherwise cause economical damage. The risk of injury or damage caused by the surgical robot, due to internal failure or deterioration is called safety for both patients or anyone else surrounding the robot. If the surgical robot is compromised or otherwise is hit by a cyberattack, this is a security\footnote{Cybersecurity and security are interchangeable in this context, but we use security throughout the rest of the paper.} failure. Safety failures can lead to injuries, but cyberattacks from individuals or organizations outside of the hospital are now able to cause safety failures as well.\footnote{Such as unwanted movement of tools inside the patient or the machine stopping entirely, see \cite[397]{Alemzadeh2016}.} Because surgical robots are CPS and are always connected to a network, security failures can cause safety failures. This means that cyberattacks on a surgical robot before or during operation can lead to physical injuries on the patient or the operator.

Expanding the understanding of what constitutes cyberattacks are necessary because it is only going to become more commonplace. Any action that is intentional and seeks to induce failure, is considered a cyberattack.\footnote{See working definition of all studies of adversarial behaviour/failure/attacks, eg \cite{Zeng2019}.} The term 'adversarial failure' and 'adversarial attack' is used for this paper instead, as this allows us to consider non-adversarial failures alongside it, which may lead to the same kind of injuries as an adversarial failure. Furthermore, there is a distinction between attack and failure, since the latter implies a failed state of the machine while the first can merely be an attempt to cause it (which may succeed).\footnote{Within security research, failure states of systems like a surgical robot being controlled by an adversary, are different than the attack going through but not causing a failure. For an example of how the failure state is studied and compared to the adversarial attacks (here adversarial examples), see \cite{Tomsett2018}.}

Surgical robots present unique security considerations, and this paper will therefore include a framework for considering specific ways that a surgical robot can fail and potentially harm the patient. These CPS specific risks will be analysed from a legal perspective as well,\footnote{Both in this paper and hopefully by many others in the future.} since this will show whether existing systems are capable of handling and otherwise mitigating them, both concerning security (both before deployment and after) but also the legal aftermath when the injuries have occurred.\footnote{Whether legislation must updated as frequently as the technology regulates, is not the scope of this paper, but it seems to be an open question regardless of the papers in the area, see eg \cite{Koops2006}, \cite{Ohm2010}, \cite{Reed2007}.}

Because the EU presents both a huge market with relatively homogenised rules, as well as somewhat digitised member states, we focus on their legislation regarding surgical robots.

Surgical robots are without a doubt considered medical devices, which means they are regulated by the European Medical Device Regulation (MDR),\footnote{Regulation (EU) 2017/745 of the European Parliament and of the Council of 5 April 2017 on medical devices, amending Directive 2001/83/EC, Regulation (EC) No 178/2002 and Regulation (EC) No 1223/2009 and repealing Council Directives 90/385/EEC and 93/42/EEC [2017] OJ L117/1. The MDR has applied fully since 26 May 2021.} since they fit the definition of ``medical devices for human use'' seen in Art 1(1). In the future, AI may even be able to partially or fully control surgical robots, which only make the role of security exponentially larger.

Any discussion involving EU law means national law in each member state apply to them as well, because the use of surgical robots will always be covered by national healthcare rules and EU standards.\footnote{The European Commission is aware of this, see eg 'Report on the safety and liability implications of Artificial Intelligence, the Internet of Things and robotics', \url{<https://eur-lex.europa.eu/legal-content/EN/TXT/PDF/?uri=CELEX:52020DC0064&from=en>}, last accessed 14 December 2021. However, they do not go into a national legal system discussion.}

Therefore, because patients who are injured by surgical robots caused by adversarial failures would want to recover damages by legal means, we present an example of how this would come to fruition in a specific national legal system in the EU. Regulation and law in general can be tools to mitigate and otherwise regulate risks. We therefore use both national and EU law, to the issues that adversarial attacks on surgical robots that lead to injuries present. 

This gives us an in-depth perspective on when the attack is successful and someone has to cover the damages done to the patient\footnote{We do not discuss the situation where the attacker is identified, since this would devolve into general criminal prosecution and the subsequent civil lawsuits.}, and the obligations and liabilities incumbent to the manufacturers in the context of the MDR. Furthermore, the possible actions by regulators that can be directed to manufacturers are discussed as well, as these will affect the manufacturer and maybe even later lawsuits against them.

The structure of the paper is as follows.
We first include novel security considerations that need to be stated when it comes to surgical robots (2), then set the scene for the use of legal sources (3), which leads into an analysis of the MDR (4), that primarily focuses on liabilities and obligations of manufacturers with some comments on guidance and accessories\footnote{Since these present novel problems.}, the main powers regulators have over manufacturers, and also comments on the current draft of the EU's proposed future AI regulation in relation to surgical robots. We then go through an in-depth analysis of the situation where an adversarial attack on a surgical robot causes damage to the patient, and how the patient would be able to get compensation for it in Denmark, here through civil lawsuits and a public compensation measure (5). Finally, we discuss our findings with a focus on novel details (6), necessary future work (7), and the final concluding remarks (8). 

\section{Definitions}
Before we initiate any sort of legal analysis, we need to define certain security terms and concepts. In this section we define what surgical robots are, how they are considered cyberphysical systems, what adversarial and non-adversarial failures are, and which adversaries would try to induce these failures. These terms enable us to understand and eventually tie law and security together.

\subsection{Surgical Robots}

The two terms active surgical robotics and telerobotics are closely related and generally explain what we see as surgical robots. Active surgical robotics\footnote{An example of the first generation of actual robotic surgical systems, could be the experimental PUMA 200 manipulator from 1988, which would define entry orientation and location of a surgical needle. The operator would then insert the needle as defined by the robot, see \cite[2]{Bergeles2014}.} means robots with pre-programmed data and computer-generated algorithms that function without real-time operator input.\footcite[114]{Hockstein2007} While also containing these features, telerobotics additionally emphasizes a remote control of a robot by a human. Control of the robot can be completely manual, or supervisory, the latter requiring substantial intelligence and/or autonomy for the robot.\footcite[742]{Niemeyer2008} A telerobotic system has an operator-site and a remote-site. The operator-site usually has an acoustic display, a visual display, a tactile display and a haptic display. Remote-site usually has acoustic/visual haptic/kinesthetic-tactile sensors or actuators. For this paper, we focus on telerobotics, which we will merely call surgical robots, since they currently all require a network connection to function properly.

\subsection{Cyberphysical Systems}

Surgical robots are cyberphysical systems (CPS), which means they seamlessly integrate computation and physical components into their operation.\footcite[2]{Solicitation2014} A generic description would be: the lowest level starts with sensors and actuators, which are connected to a field or a sensor network, all of which would be managed by a control system, that itself would be bound to a control system network.\footcite{Kobara2016} 

The human machine interface would exist at this level, which would be where the operator of a surgical robot would reside. Autonomous cars, smart grids and IoT devices all fit these criteria, but with these features come many vulnerabilities, which are included with the connection to a network or the internet. As they have physical components, these devices can interact and affect the health of humans, and so can an adversary that successfully attacks the system. This shows that surgical robots, in this context, can be considered safety-critical systems, because of the potential risks imposed on the patient.\footcite[15]{Alemzadeh2016b} CPS consists of many hardware and software systems combined, and each can be manipulated from the outside, even if it is loosely isolated from the internet.\footcite[788]{Kobara2016} Essentially, we assume that surgical robots are CPS, which is used to connect it and this paper to the wider field of CPS as some of the legal considerations can apply to other systems which interface with humans in the same manner.

\subsection{Adversarial failures}

Given the extensive developments in CPS, we can assume it is hard for manufacturers, doctors, hospitals, robot operators, engineers, lawyers, and policymakers to keep up with the developments in the world of these systems. As these solutions become popular, there is ever greater need to understand how they fail, since this is fundamental to any litigation and assignment of responsibility.

Adversarial failures are caused by an active adversary who attempts to induce failures to attain their goals, such as manipulating the surgeon's commands, inferring the surgical procedure to compromise patient privacy, or the infringement of intellectual property such as the robot's algorithm or trade-sensitive data like the surgeon's inputs. 

There exists other taxonomies for security\footnote{See some of them in the following 5 examples: \cite{Rizvi2018}, \cite{Quarta2017}, \cite{Papp2015}, \cite{Aslam1996} \cite{Carl1994}.} and safety,\footcite{Vasic2013} but we feel that a separate taxonomy is needed due to the specific issues that surgical robots pose.\footnote{We agree that there are close similarities between IoT and surgical robots, but not enough to warrant using the same taxonomies. Specialised types of CPS require their own considerations.} The taxonomy is also highly relevant for the legal analysis further on in this paper. Our proposed categories of adversarial failures for surgical robots are as follows:

\begin{enumerate}
    \item \emph{Manipulation Attacks.}\footnote{The word attack is used here, regardless of whether it is a failure for historical reasons.} The adversary covertly modifies the instructions to get a different desired response. This is understood in the broadest sense, since it can be initiated in any part of the CPS that surgical robots consist of. \footnote{This was tested in practice on the RAVEN II open platform, which is similar to current surgical robots, see \cite{Alemzadeh2016}.} The attacks were injections of unintended user inputs, or motor torque commands, which required access to the master console or control software. The effect of this failure would be unintended jumps, movements or for the robot to completely stop.

    \item \emph{Subverting robotic control.} The adversary hijacks or otherwise makes changes in the robot's control. 
    
    This is different from manipulation, since this can be done on the network the robot receives signals from, and focuses on the control, not manipulating existing actions. A practical test worth mentioning, done on the same surgical robotic platform as before,\footcite{Bonaci2015a} where packets were delayed or changed between the operator and robot, and using this technique, they were also able to hijack the surgical robot.\footnote{This was done by fooling the robot to believe that input believing packet loss was occurring, but not long enough to interrupt the operation, and the surgical robot would then only be able to be controlled by the packets sent by the adversary.} While the first two can only cause delays in movements, the latter completely enables the adversary to do as they please, potentially harming the patient.

    \item \emph{Reprogramming the robot.} The type of access that is needed to manipulate the robot, may also allow access to change the software as well. The failure consists of changes in software on any level, and while we do not have practical examples for surgical robots, the severity of such failures on the patient or operation in general is large enough to raise concern. The possible enabling of other failures or other newly programmed actions are great, and this only shows how important maintenance and routine validation of equipment is.

    \item \emph{Misappropriation of trade secrets.} This is seen as the attackers recreating the underlying technique of surgery by collecting surgical control instructions over time. This failure can be initiated over the network or inside the surgical robot. While it cannot harm the patient, it is misappropriation of the techniques used by surgeons, and could in the future constitute the basis for data sets that AI or machine learning algorithms can use to replace the operator. Collecting this without the consent of the surgeon is both unethical and likely violates rules on trade secrets and data protection.

    \item \emph{Poisoning the feedback loop.} The adversary covertly modifies the camera and/or other sensory outputs sent to the surgeon. Sensory inputs are currently vital for showing where the procedure inside the patient is at, as well as what the surgeon is currently doing. If any of these are changed, the risk of injury of the patient increases. The difficulty of resetting or returning the robot to its initial position is further hampered by any feedback being off or wrong, which makes this a dangerous failure.

    \item \emph{Software vulnerabilities.} Any vulnerability that an adversary can make use of to commit further attacks on, is considered a failure as such. It is also the broadest, since it covers any part of the surgical robot and its accessories. Unlikely the failures above, this is passive and not necessarily caused by the adversary, but instead enables them to cause failures because of it. 
    
\end{enumerate}

\begin{figure}[h]
\caption{Illustration of adversarial failures for surgical robots, which includes what is compromised, here integrity, confidentiality and availability.}
\centering
\includegraphics[width=\textwidth]{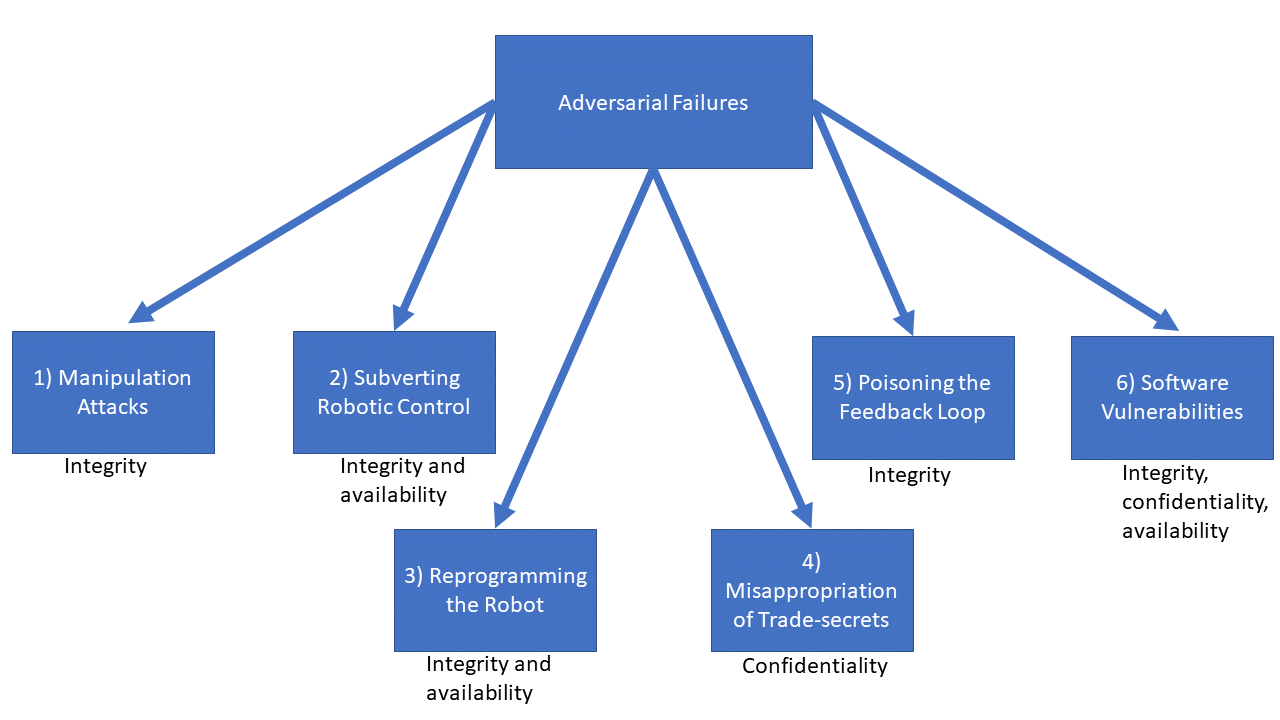}
\end{figure}

\newpage
    
\subsubsection{Non-adversarial failures}

We do not intimately discuss non-adversarial failures in this paper, but we choose to list them for completeness. Non-adversarial failures are caused by the correct operation of the robots as per the specification, but where an unsafe outcome is caused nonetheless. We see them as:

\begin{enumerate}

\item The robot works in unintended ways because of failures in motor calibration or sensory defects.

\item The robot causes a denial of service on itself whilst legitimately trying to accomplish the assigned task.

\item The robot has an incremental bias which creeps in due to shifts in belt tensions, gear wear-and-tear and other electro-mechanical reasons.

\item The robot fails to handle shifts in lighting, shadows, tilt of surface level, noise, mist or other environmental noise in the visual or acoustic plane.

\item The robot fails to perform due to inability to function in poor network conditions or being operated in network conditions (jitter, throughput, and bandwidth) that are quite different from what it was tested on. 

\end{enumerate}

Adversarial failures can manifest via non-adversarial pathways. An attacker may manipulate neighbouring devices that are not connected to the robot via a computer network but nonetheless provides interaction pathways. For instance, an attacker may introduce subtle changes in lighting via a compromised IoT lightbulb inducing a failure in the surgical robot's image recognition component potentially leading to patient injury. This is a example of a situation where a safety failure in the robot is induced via a compromised device in the vicinity of the said robot. In security literature, these are referred to as stepping-stone attacks, where the attack is carried out through indirect influence rather than direct engagement between the surgical robot and the attacker.\footnote{See eg \cite{Nicol2014}.} 

This approach affords relative anonymity to the attacker and makes attribution harder as the attacker is separated several steps away from the intended target due to the indirect nature of engagement.
 
\subsection{Adversarial model}

The identity of the adversary, the party that seeks to cause adversarial failures, have different priorities and focuses, so we decide to assume certain things about them, as adversaries are traditionally modelled in security.

Our adversarial model includes cybercriminals, disgruntled employees, terrorists/activists/organized criminal groups, and nation states,\footcite[1 - 2]{Alvaro2009} as well as competing surgical robot manufacturers. We choose to have the widest range of actors possible, since the choice of them and which failure they want to induce, can change the outcome of the analysis in the following sections.

\begin{figure}[h]
\caption{Illustration of which adversarial failures the adversaries induce.}
\centering
\includegraphics[width=\textwidth]{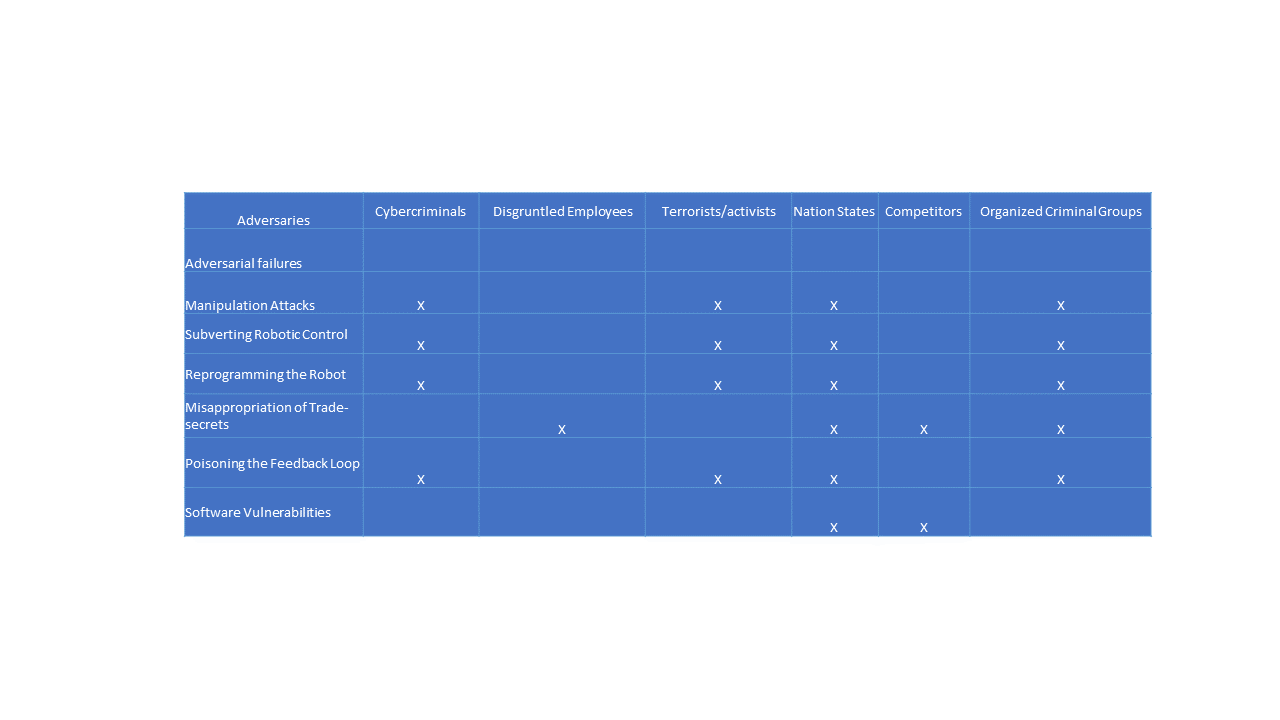}
\end{figure}

We assume that stronger players can induce many failures, while competitors and disgruntled employees would only induce a few. 

We leave out any attacks that can cause injury from those two, since none of them would have the intention to cause them in the first place. Cybercriminals and organized criminals groups overlap, and can generally both create all failures except for software vulnerabilities, but we assume that misappropriation of trade secrets would be done only by the organized party. Terrorists/activists would want to create as much of as much disruption as possible, which is why they would go for failures that cause this. And nation states are capable of everything, with the highest amount of resources at their disposal.

\section{Legal Sources}

Surgical robots are used and sold both at a national and EU level, which means that both legal spheres apply to them at the same time. As indicated in section 2, we are interested in analysing the very real situation where surgical robots are attacked by an adversary, and which legal tools the victims as well as the manufacturers and other parties can make use of. The EU rules include the MDR and relevant guidance.\footnote{There exists further relevant legislation regarding both data processing and security, where the NIS Directive, Directive (EU) 2016/745 of the European Parliament and of the Council of 6 July 2016 concerning measures for a high common level of security of network and information systems across the Union, OJ L194/1, is hugely relevant. This is because operators of surgical robots will be covered by it in Art 5(2) and its Annex I, 6. Depending on the implementation of the directive, this leaves an additional layer of security that must be fulfilled by any operator of a surgical robot.} We will also include some long term perspectives from the current draft of the EU's proposed AI regulation, which is not current law, but will be eventually in some capacity.\footnote{Proposal for a Regulation of the European Parliament and of the Council laying Down Harmonised Rules on Artificial Intelligence (Artificial Intelligence Act) and Amending Certain Union Legislative Acts, \url{<https://eur-lex.europa.eu/legal-content/EN/TXT/?uri=CELEX:52021PC0206>}, last accessed 14 December 2021.}

The Danish legislation includes rules for suing manufacturers, as well as law for public authorities since one approach to compensation for damages is through an administrative system.

\section{The European Medical Device Regulation}

The injured patient will be glad to know that the surgical robot is initially regulated by the MDR, which gives them some rights, but more importantly places obligations on manufacturers which may prove important in a future lawsuit or Patient Compensation case. For the sake of completeness, we give a full overview of the MDR and how it relates to surgical robots and security, including obligations of authorities and manufacturers, while we refrain from doing a full analysis on its structure and how it is different from the past Medical Device Directive.\footnote{Council Directive 93/42/EEC of 14 June 1993 concerning medical devices [1993] OJ L169/1.}

\subsection{Background}

The MDR is designed to achieve a balance between a high level protection of health for patients and users, as well as high standards for quality and safety of medical devices.\footnote{See preamble 2 in the regulation.} This fits the scope of the regulation, stated in Article 1(1),\footnote{For further details on the interpretation of Article 1(1), see \cite{Ludvigsen2021}.} which is a focus on laying down rules for placing medical devices on the market.\footnote{Contrary to a directive, a regulation is a binding legal instrument, that is directly enforceable by the states and the EU, see Art 288 in the TEU.} It there is any doubt as to whether surgical robots are included within the regulation, we note that Article 1(1) includes any “medical devices for human use”. This encompasses surgical robots. Additionally, surgical robots are not included in Article 1(6). This article constitutes the negative definitions (exclusions).

\subsubsection{Accessories}

Like other robots and CPS, surgical robots make use of software and physical additions that on a practical level will be accessories. But there are certain requirements for them to be considered accessories in regime of the MDR. The reason why identification of these is important, both to a manufacturer and to a potential injured patient, is because many of these could be points of entry for adversarial attacks, or be the actuators that injure the individual or the operator.\footnote{This section is included for completeness, but must be explored further elsewhere, see section 7 of this paper.}

Accessories of surgical robots are governed by the same rules as the robots they are used with,\footnote{See Art 1(1).} even if they do not attain the status of medical devices. We must define what accessories are in this context as stated in Article 2(2),\footnote{We exclude any tools that are themselves already medical devices, such as the scalpels and other tools used by surgical robots, which will be covered by Art 2(1) just like the robot itself, and even if they practically are accessories, they are not included in the legal definition here or in the MDR.} because a surgical robot as a system has more accessories than most medical devices, which puts it in a unique position in terms of liability for the manufacturer

Any traditional accessory that is not a medical device, such as sensors for surgical robots, are included in Article 2(2), first definition. But the second definition expands and includes anything that exists “to specifically and directly assist in the medical functionality of the medical device(s)”. To be considered an accessory, it therefore has to also specifically and directly assist with the medical functionality of the surgical robot. For telerobotic surgery this is both the encrypted connection, the local network that enables it and the operator screen and equipment that controls it elsewhere. This is therefore an expansion of what an accessories means in a medical device context.

\begin{figure}[h]
\caption{Example of accessories of a surgical robot.}
\centering
\includegraphics[width=\textwidth]{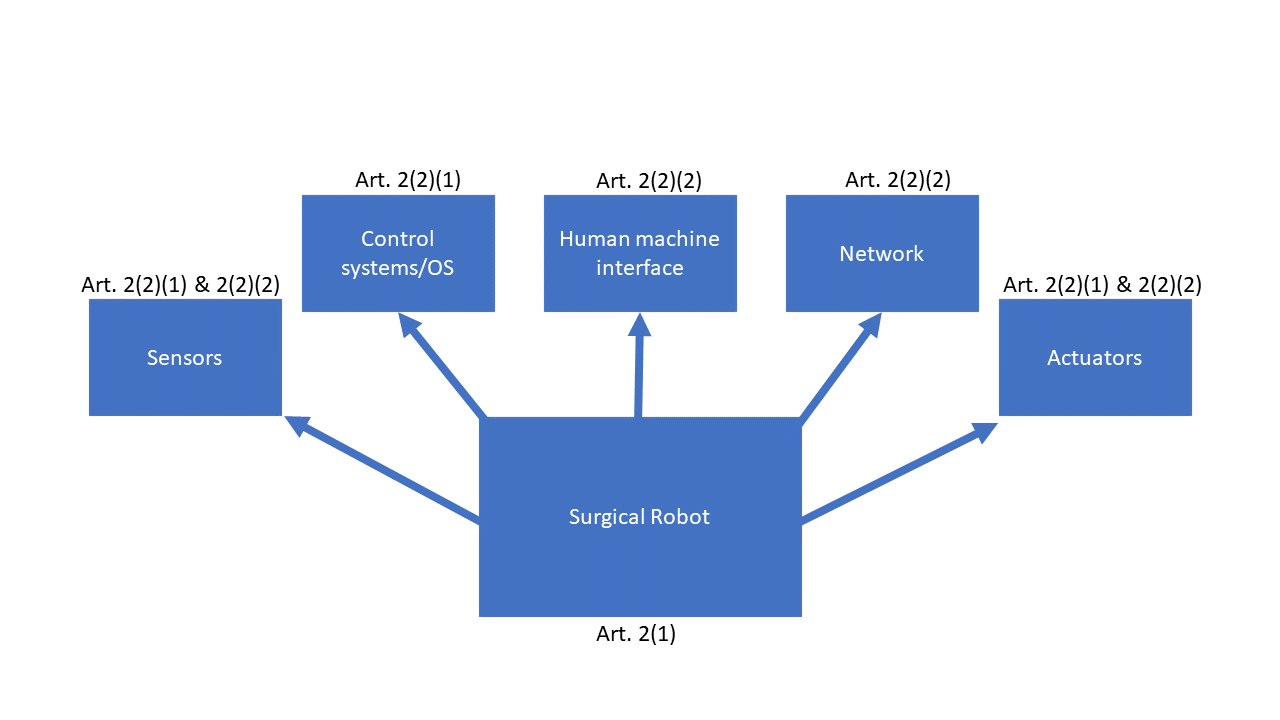}
\end{figure}

\newpage

This does not mean that the accessories, if not directly a part of the surgical robot, make the manufacturer specifically responsible. It is likely that the operator or end user must maintain and keep them updated,\footnote{Depending on the contract between them.} and a separate evaluation on whether they would be considered medical devices is taken by the member state.\footnote{See preamble 8.} The relevant national regulatory authority is that of the accessory manufacturer's place of business, as is the case later in this paper. But if the accessories, such as specific equipment for the physical part of the surgery, are a direct part of the robot, and is not included in the exception in preamble 19 of the MDR,\footcite[6]{Ludvigsen2021} the manufacturer may be responsible for any safety or security issues they present. This expansion of the concept of accessories regarding surgical robots increases the amount of possible targets for lawsuits from the patient, since software or actuators that are accessories but which may have initially suffered an adversarial failure to allow access into the surgical robot, will bear their own liability alongside the manufacturer of the surgical robot.\footnote{The scope of this paper does not allow these observations to be further discussed, for more details see section 7.}

At its core, accessories will have their own known adversarial/non-adversarial failures, which the manufacturer may responsible for either through Article 2(2)(1), non-medical devices which enable the medical device, or Article 2(2)(2), non-medical devices who specifically/directly assist a medical device. From this we create 5 categories of specific accessories for surgical robots, which combine the real complexity which is surgical robots as CPS with the flexibility of the MDR, to allow future authors to explore the very thin veil between the manufacturer of the entire CPS being liable, versus when the manufacturer of the accessory is.

\subsection{Guidance on Cybersecurity}

Like national law, EU law has additional documentation and guidance that can be used by different parties affected by it. One of these the is
``Guidance on Cybersecurity for medical devices'',\footnote{For a separate commentary on this, see \cite{Biasin2020a}.} which is issued by the Medical Device Coordination Group (MDCG).\footnote{Established by Art 103 in this regulation.} In general guidance can be seen as legally binding,\footnote{This is prevalent in eg Danish law, if the guidance is purely made for a specific public authority, but can be problematic to always impose in EU-law, as it relies on national authorities and interpretations entirely.} a tool to guide interpretation or at least act as guidelines for the parties.\footnote{But because it is guidance, it cannot be taken as a positivist legal means to regulate security.} 

The MDCG, while having created this, does not issue legally binding guidance, as there is nothing stating this in the regulation, but in the future Article 103(8) does allow them to create recommendations or opinions in emergencies. For this reason, we have to view the guidance on security as non-binding soft law.

While the MDR does not explicitly consider the safety to security problems, the guidance does on page 10. It equates security risks having a safety impact, which here for us would refer to damage to a patient caused by an adversarial failure. It argues that because of this, Annex I\footnote{Defines further requirements for medical devices, see section 4.3 in this paper for more details.} has to both be interpreted in a safety as well as a security manner. This dispels any doubt whether the MDR can be used to argue for lawsuits on the basis of adversarial failures.

\subsection{Manufacturers}

Surgical robots can be put on the market by several different parties,\footnote{Such as importers and distributors, see Art 13 and 14 of the MDR.} and as mentioned earlier, we have chosen the manufacturer as the most important one due to their central role in the surgical robot’s life cycle.

As a general rule, the manufacturer answers only to the regulator in its place of business.\footnote{See Art 10(14).} However, any patient in any member state can sue any manufacturer, because of the rule of special jurisdiction.\footnote{See Art 7 in Regulation (EU) 1215/2012 of the European Parliament and of the Council of 12 December 2012 on jurisdiction and the recognition and enforcement of judgements in civil and commercial matters [2012] OJ L 351/1.} 
\newline

The relevant obligations are the following:
\begin{enumerate}
\item The system of risk management (Article 10(2)).
\item The system for quality management (Article 10(9)).
\item Sole responsibility for devices (Article 10(1)(12)(13)(14)).
\item A system for financial responsibility (Article 10(16)).
\item Annex I specific obligations.
\end{enumerate}

\paragraph{The system of risk management.}\footnote{This is further defined in Annex I, section 3.} The regulation defines it as  a continuous iterative process through the surgical robot’s entire life cycle. The system has to identify and analyse all foreseeable hazards, estimate and evaluate risks associated with/occurring during intended use and the future, eliminate or control those and evaluate this information and combine it with the data gathered from the post-market surveillance system.\footnote{As seen in Art 83, which requires manufacturers to have a system in place for surveillance of the post-market situation of their device, be it academic or technical data.}

The term hazard does not literally include adversarial failures, but since they can cause hazards in a safety manner, like the surgical robot jumping or possible getting hijacked which risks the patient and anyone nearby, they should naturally be included. Misappropriation of trade secrets or software vulnerabilities cannot directly cause physical harm, and should therefore not be included, unless they clearly cause further adversarial failures. From this, all other adversarial failures must be eliminated or controlled.

The security guidance summarizes these parts of the system as a circular information flow, consisting of a risk management plan, assessment, risk control, evaluation, review(s) before release and post-market activities.\footnote{See p. 17 of the Guidance.} But this is not directly derived from the regulation.

\paragraph{The system of quality management} While this system includes the risk management above, it also has others elements. Firstly, identification of applicable general safety and performance requirements/exploration of options to address it. This is mandated,\footnote{See Art 10(9).} and must be addressed separately from the rest. Secondly, the post-market surveillance system, which in security terms must identify all incidents that can occur to the surgical robot in the future. As both angles must imagine all types of safety failures, any kind of security failure, by an adversary or otherwise, must clearly be included.

\paragraph{Sole responsibility} The term sole responsibility refers to the manufacturer’s role as both the creator and controller in Article 10(12), in the sense that any non-compliance by the device has to be relayed to the regulator, and as a partner with the regulator, since they have to cooperate and follow requests given in Article 10(14). This refers back to Article 10(1), which solely states that the devices should be designed and manufactured to comply with the regulation. Generic software, other than it being being updated, is not included, but accessories are. This is a clear way to show liability in national law in the case of doubt, as it constitutes a legal rule which will be breached if the surgical robot suffers an adversarial failure that could have been prevented or mitigated.\footnote{We come back to this specific liability measure in section 5.3.1 of this paper.}

\paragraph{A system for financial responsibility} This is defined in Article 10(16), where compensation schemes specific to each country are mentioned, which usually involves initial insurance coverage, but also product liability lawsuits and national law. The risk class, type and size of the manufacturer plays a role in which measures, like insurance, that they have to undertake, but the article takes national protective measures into account – which for example includes the Patient Compensation from earlier. This shows that the regulation leaves all legal remedies and considerations concerning litigation up to the member states and insurance solutions, which detracts from its value as a regulation, because it reduces the effect of the proposed harmonisation.

\paragraph{Annex I} This annex further defines requirements for the medical devices, and according to the guidance and if read literally, section 17 on electronic programmable systems should be the focus when it comes to surgical robots. Section 17.1 requires that the devices are designed for repeatability, reliability and performance. If a single “fault” is found, it has to be eliminated or reduced as much as possible. Whether fault only refers to non-adversarial failures or the opposite is unclear, but considering the guidance’s emphasis on this part, an interpretation that sees it as adversarial failures seems appropriate. 

This is supported by safety faults being the focus elsewhere.\footnote{See eg Chapter III, Annex I.} Since these requirements are not part of the risk management system, this further emphasizes that preventing any adversarial failure besides misappropriation of trade secrets, and repetition of these requirements in different ways cements its importance in the production, sale and usage of surgical robots.

Section 17.2 requires the software used in devices to be developed and/or manufactured with the “state of the art”. State of the art is used sparingly in the regulation, but has not been included in any of the central articles. The term in section 17.2 equates to regular updates and maintenance of the software, and it has to consider the life cycle of the device and \emph{information security, verification and validation} of the robot. The three last categories would imply that it should catch all adversarial failures, with security preventing manipulation attacks and subversion of robotic control and perhaps misappropriation of trade secrets, and verification catching reprogramming of the robot and poisoning of the feedback loop, and validation reinforcing whether the security is adequate or not. State of the art would then prevent software vulnerabilities by regularly identifying and erasing them. However, state of the art only requires what term encompasses, which also means that anything that the industry does not know or what is not expected of it, it does not require the manufacturer to do. This includes which adversaries that should be defended against, with nation states being impossible to include because of their immense power. There are certain parallels with how we will discuss ``current knowledge'' of an industry in section 5.1.1, and in practice the legal standards will likely overlap.

Section 17.4 requires that the manufacturers decide on minimum requirements for hardware/IT networks characteristics/IT security measures, that allows the software to run ``as intended''. This allows the manufacturer to technically set standards that could be problematic in the long run, since it might not prevent more complicated and dangerous adversarial failures.

\subsection{Regulators}

The other subject, the regulator, is defined as the competent authority. Member states designate these themselves.\footnote{See Article 101.} As with manufacturers, we pick out a range of rights of the regulators which are relevant when considering surgical robots and which will affect the responsibility and liability of the manufacturers.

\begin{enumerate}
\item Right to request documentation and punish the manufacturer if they do not cooperate (Article 10(14)).
\item Market surveillance activities (Article 93).
\item Evaluation of devices suspected of presenting an unacceptable risk or other non-compliance (Article 94).
\item Procedure for dealing with devices presenting an unacceptable risk to health and safety (Article 95).
\item Other non-compliance (Article 97).
\end{enumerate}

\paragraph{Right to request and punish.} This part of the article contains the special right for the patient in its paragraph 3, but we focus on 1 and 2. In paragraph 1, the manufacturer must provide documentation to demonstrate conformity of the device, or samples free of charge or access to the device. Further, they have to cooperate on any corrective action to eliminate or reduce risk for devices they put on the market.

If this is in some way not possible, the regulator has the right in paragraph 2 to take all appropriate measures to prohibit/restrict/withdraw/recall the device.\footnote{See Art 10(14), second paragraph.} The right is not built up as an immediate use of force, but rather the opposite. It is instead based on trust in the manufacturer fulfilling the requests from the regulator dutifully. 

It is not specified whether the regulator has the necessary knowledge or personnel to request actions or documentation that relates to security, but because of the existence of the guidance this might be the intention. While this is not a liability for the manufacturer, it constitutes a risk of their product being banned or forcefully withdrawn or changed - all of which are increased costs.

\paragraph{Market surveillance activities.} This activity resembles what most are familiar with from national food regulation authorities.\footnote{See eg Chapter II in Regulation (EC) No 853/2004 of the European Parliament and of the Council of 29 April 2004 laying down specific hygiene rules for on the hygiene of foodstuffs [2004] OJ L139.} Review of documentation, physical or laboratory checks are possible, as is requesting documentation from other parties than the manufacturers and unannounced inspections.\footnote{See Art 93.} How this can be applied to security cannot be literally read, but considering the wide power the regulator has, it is theoretically able to thoroughly review and inspect risks that might lead to adversarial failures.\footnote{This is both promising, and at same time most likely a huge weakness of the MDR. We see that in the EU AI draft reg., that there is room to require specific staff and expertise from the type of national authorities needed for AI, and this should probably be spread to all types of regulation. See section 7 of this paper for more.}

\paragraph{Evaluation of non-compliance.} If the regulator takes notice of there being an unacceptable risk to the health/safety of patients/others, or if the device seems to not comply in general, they are then allowed to carry out a more thorough investigation that includes the complete check of compliance of the regulation.\footnote{See Art 94.} It is unknown whether this includes penetration testing or other validation measures of the devices.\footnote{The lack of information on the practice of regulators is lacking in every aspect, and future research in this area would be greatly appreciated.}

\paragraph{Procedure for devices that risk health and safety.} If the regulators are confirmed in their suspicions, they first ask the manufacturer to take all appropriate and duly justified corrective actions to restore compliance, and until then, themselves proportionally restrict the availability of the device.\footnote{See Art 95.} This latter point means recalling the device in practice. And if this is not done, this reverts back to Article 10(14), where the regulators can forcefully remove the robot from the market.

\paragraph{Procedure for non-compliance.} If the evaluation showed other non-compliance, the regulator can react in a similar fashion to Article 95. The requirement for unacceptable risk to health etc., is not present here, but the powers are the same. This is interesting because it can potentially include adversarial failures that do not have a risk to the health and safety of anyone, for example misappropriation of trade secrets and software vulnerabilities. It remains to be seen how this can be used in regards to surgical robots.

\subsection{Comments on the MDR}

It is no secret that the Medical Device Regulation does not literally mention security, adversarial failures or even robots. This is because it is a EU regulation, which strives to encompass all possible medical devices,\footnote{See Art 1(1).} while not literally aiming to be technologically neutral. We do not criticize or comment on this, but it is also one of the most important legal frameworks for surgical robots for the foreseeable future. No one would doubt that the regulation improves and continues what the directive did.\footnote{See preamble 4.}

The regulation chooses to not have a focus on utmost prevention of adversarial failures \emph{ex ante}, and explicitly does not say this outside of the annex and vaguely in other spots. If it had done this, both the safety and security of the patient would have been the focus. However, the safety of the patient is not mentioned in Article 1, which shows that it is not in the explicit scope of the regulation. 

This does not mean that issues cannot be addressed in practice by the manufacturers and regulators, and preamble 101 leaves room for further legislation down the line to ensure the goals of the regulation, which could include rules for security specifically. Future case law at CJEU can also further address these issues. The choice of leaving many important security issues to guidance is likely deliberate. But the future of healthcare is in the field of CPS, which will have issues with adversarial failures. The decision to equate safety issues with security issues because of security in guidance instead of the regulation is not adequate considering the guidance is not legally binding. But as shown, the guidance functions well as a tool to interpret articles and the annex. And it is possible that there will be \emph{lex specialis} that manages the specific issues that the field presents later, but they are unlikely to initially be regulations, and will therefore have less weight in the world of EU-law.

\subsubsection{Considerations for manufacturers}

Earlier, \emph{we choose} five core obligations which amount to additional responsibility for the manufacturer for a surgical robot, and which can become means for increased liability in lawsuits in the future regarding adversarial attacks. Three systems, which are risk and quality management, and financial responsibility, and the last 2, which are sole responsibility for the devices and all requirements for them in Annex I (including state of the art).

Even with quality and risk management systems, defined rights for regulators and tight obligations for manufacturers, and a forced state of the art idea in place, one must doubt if the manufacturers would still prefer to let the accidents happen, and let litigation solve any issues later. Prevention is mentioned in the regulation, but if it fails to enforce prevention in specific technical, yet physically dangerous aspects, such as those CPS represent when adversarial failures occur, it ends up having to solve these issues \emph{ex post}.

However, the regulation does succeed in stating separate obligations to reinforce security, with the same systems and the requirements in Annex I cumulatively functioning together.

\subsubsection{Considerations for regulators}

For regulators, \emph{we choose} five powers which are highly relevant when considering adversarial attacks on surgical robots, and which represent real threats to the bottom line and rights of the manufacturers. These are right to request documentation (with consequences if no compliance), market surveillance, evaluation of suspected risk or other non-compliance, means to deal with unacceptable risk to health and safety, and other non-compliance. Several of these rights enable the regulator to withdraw or otherwise limit the sale of the surgical robot, but at its core, having the manufacturer willingly improve and prevent for example adversarial failures by themselves is the ideal.

As can be seen, regulators do not have to explicitly fulfil any security obligations,\footnote{Outside of maintaining and being part of the various information sharing systems.} but the question remains whether they can each sufficiently inspect and regulate the security of surgical robots. There is not a literal requirement for this in the regulation. Article 101 merely defines how that the regulator has to be designated, but there is nothing concerning specific specialized staff that can handle security. This is widely different compared to regulations such as the GDPR\footnote{Specifically chapter IV, section 4 and chapter VI.} which lays down a rigorous structure.\footnote{The EU AI regulation draft is an example too, see section 4.6 of this paper.} Instead, the MDCG are supposed to coordinate efforts across the EU, but they have not been given enough rights and tools to inspect or enforce practice in Article 105 and 106, which combined with lack of specifications for the staff of regulators\footnote{None in Art 101.} creates a worrying environment for effective enforcement of the regulation.

It is clear that the regulators rely on documentation and information gathered by the manufacturers and its partners, as seen in clinical evaluations and investigations\footnote{See Chapter VI in the regulation.} which we have not covered, but as shown they also posses several rights that in the future could be central. Inspections, requests to correct issues and the sharing of all documentation between all regulators from all possible manufacturers via the MDCG and their systems, as well as forcefully moving dangerous medical devices off the market by any means, show that the system and its players may solve some of the issues we have shown with the regulation. This includes the special right for patients in Article 10(14), mentioned earlier, that enables them to make use of the documentation gathered by the regulators as well.

There is an additional issue with the regulators which is shared among other types of product legislation in the EU: what should a patient do if they suspect that a medical device which was used on them violates the MDR? The first step is to alert the regulator, which would then lead to any number of the procedures above, but if they choose to not react, civil litigation against the manufacturer is possible within some jurisdictions, as would lawsuits against the state for not enforcing the MDR in this situation.

\subsection{The Future AI Regulation}

Regulation of AI on a EU level is still far away as of the time of writing\footnote{Civil Liability and AI have been considered in 'Artificial Intelligence and Civil Liability', commissioned by the European Parliament's Committee on Legal Affairs, \url{<https://www.europarl.europa.eu/RegData/etudes/STUD/2020/621926/IPOL_STU(2020)621926_EN.pdf>}, last accessed 14 December 2021. The report has had no real legal consequences as of yet.} but the European Commission did publish its proposal for a draft for one the 21 April 2021. 

Surgical robots may in the future or have already made use of AI,\footnote{Ibid, p. 111 - 113.} which means it is covered by this regulation in the future. We can derive this from the draft in Articles 1(a), 3(1),\footnote{All current surgical robots will be covered by Annex I, b, and in the future most likely a and c.} and we can quickly see that the AI they use will automatically be high-risk in Article 6(1)(a), because they will be used on medical devices which is explicitly mentioned in Annex II(11). 

There are no additional rights for patients in the draft regulation. 

They are as a whole excluded and are not included in its purpose.\footnote{See Art 1 and 2, that do not mention patients or data subjects.} This is due to the role which this regulation will play. Like the MDR, this is a piece of legislation aimed at EU product requirements, which is clear with the future amendments to relevant product legislation where AI will appear.

Unlike the MDR, security and adversarial attacks are considered,\footnote{See Art 14 and 15.} and the regulation will require the staff and regulators as such to possess certain qualities.\footnote{See Art 59.} This is very promising for all fields where safety becomes a security matter, and expanding this approach to all product regulation in the EU would be very beneficial, as the requirements and rigorous and future proof (if followed loyally).

\section{Danish law}

In this section, we show how patients can use applicable law to recover their damages from an adversarial attack on a surgical robot in Danish Law. We show whether a lawsuit with product liability, reimbursement outside of contract or a case in the Patient Compensation system in Danish law will be likely to succeed, and we do so from the perspective of the patient as the claimant and the manufacturer as the defendant. We analyse these means as they are the only way in which an injured individual could claim compensation from the damage caused on them, and because manufacturers of such robots must be aware of their liability and the following potential risks of litigation.

\paragraph{Initial comments}

Robots or CPS as such in Danish law do not have \emph{lex specialis} made for them, and outside the implementation of the NIS directive, security and adversarial attacks do not have any either.\footnote{Directives are implemented into national law, not directly used like regulations, see Art 288 in the Consolidated Version of the Treaty on the Functioning of the European Union.}

Before one makes use of the most general approach to compensation, two other types must be considered, as these are \emph{lex specialis}, albeit not for adversarial attacks or security. We therefore go through product liability,\footnote{Council Directive
of 25 July 1985 on the approximation of the laws, regulations and administrative provisions of
the Member States concerning liability for defective products, OJ L210/29. Implementation of the Product Liability Directive in Danish law is done through the Law of product liability, LBK nr. 261, 20/03/2007.} then Patient Compensation Association below. Patient Compensation is not a lawsuit, but a separate administrative means to claim compensation.

However, both specialized approaches build on the thoughts from reimbursement outside of contract, a case law based means to receive compensation in civil litigation.\footnote{This is not to be confused with extracontractual liability, which as a principle applies strictly to reimbursement at all times when there is no contract dictating the terms. In contrast, reimbursement outside of contract includes legal principles that are far older than the EU, which modify and set boundaries together with Danish contract law in general. Furthermore, extracontractual liability does not include the process and other rules regarding issues beyond liability, which reimbursement outside of contract always does.} Because of its special role and much longer existence than the other two, a patient can always fall back on this. To sue on the basis of reimbursement outside of contract requires that 4 specific case law based criteria for reimbursement are fulfilled: someone who is liable, quantifiable damage, a link between the responsible and the damage and that the link is adequate,\footcite[23]{Eyben2013} and these must be fulfilled cumulatively. The defendant may have acted carelessly, and the adversarial failure may have caused damage, and a link between the two can be made likely, but if the link is not adequate, the case will be ruled in favour of the defendant. We will return to this last approach after the other two. Both the lawsuits will be part of civil litigation, which initially leaves the patient at a disadvantage in terms of providing evidence.

\subsection{Product liability}

If a product is defective and causes damage, a lawsuit on the basis of product liability can be initiated. Defect is defined as the product being less safe than a person is entitled to expect.\footnote{See Art 1 in the Product Liability Directive, or § 5 in the implementation law.} Normally, this would not apply to a product (here a surgical robot) which has been purchased by someone else than the patient who was injured by it, but we have a case that allows this in Danish law.

If a patient being treated by a medical device that fails due to a defect and gets injured, we know that the patient is entitled to directly sue the manufacturer instead of the hospital in Danish law. This was answered more than forty years ago in the case U.1960.576H,\footnote{Notation for Danish case law.} where two patients kept the manufacturer of oxygen machines liable and had to compensate for the damage that was caused on them as they were hospitalized.

Product liability can be sought in three ways in Danish law. The first is on the basis of the product liability directive, the second is through a case law based approach that existed before the directive,\footnote{Also called 'delict based product liability'.} and the third is the oldest and is product liability through contract.

\subsubsection{Product liability directive lawsuit}

Lawsuits for product liability are usually initiated on the basis of the Product Liability Directive as transposed in Danish law.\footcite[498]{Obligation2015} 

First, we see that the surgical robot and accessories are included by the law, since it is a product.\footnote{See § 3 of the implementation law.} Secondly, we must consider whether the manufacturer is exempted from responsibility. Surgical robots that are not considered goods are exempted. This refers to idiosyncratic surgical robots that cannot be moved without destroying them.\footnote{This also means they are not products because they have not been put into circulation.} 

The defendant can further argue that if the surgical robot has not been put into circulation, has been designed to be used by a single hospital, or that due to the state of the art the time it was impossible to discover the defect, they are not responsible.\footnote{See § 7, part 1.} The last factor is especially important since this enables the defendant to argue against any adversarial failures caused by new or unusual means, but the burden of proof for this is incredibly high, as it is the knowledge of the \emph{entire industry}, not just what the single manufacturer knew at the time. This then becomes a case of inviting the best expert witnesses or hoping that the claimant ignores new or extraordinary research.

If the manufacturer is liable, we can continue of the rest process. The claimant must then prove that a defect exists.\footnote{See § 6.} They also have to prove the damage and the link between the defect and the damage.\footnote{The criteria are partially derived from the aforementioned reimbursement outside of contract.} 


A defect is defined as the product being less safe than a person is entitled to expect,\footnote{See § 5.} and the patient can claim that they can expect for the surgery to only fail due to mistakes by the operator or mechanical or safety failures, not those caused by adversaries. Three considerations can modify this assessment, which are: the marketing of the product, its intended and expected use and the time at which it was put into circulation.\footnote{See § 5, part 1.}

Marketing is irrelevant to the claimant, but intended and expected use will include some unusual considerations when it comes to the last exemption in § 7, part 1, 4 or part 2, which is consequences that could not be foreseen at the time of sale. Since the use of such a robot naturally will include maintenance of any level of software, evading responsibility is impossible on those grounds when considering security aspects, unless it is impossible to defend against, such as future adversarial failures caused by quantum computing. 

In our view, the exception will only apply to extraordinary adversarial attacks, as zero-day attacks and exploits are not unforeseeable and will always have a chance of occurring. Identifying the defect is crucial, which would require that the claimant obtains proof of the three adversarial failures that potentially can cause an injury on the patient. These are manipulation attacks, subversion of robotic control and poisoning of the feedback loop.\footnote{See section 2.3 of this paper.} The claimant can then require original design documentation, which an expert witness could question as to whether the surgical robot is under the risk of specific manipulation of the robot or subversion of the control of the robot over the network, as well as poisoning of the feedback loop given from visual and haptic sources.

Another approach, which is even more appropriate, is the argument \emph{res ipse loquitur}, proving that the injury was not caused by human or other error. This will force the defendant to either argue that it was caused by a safety defect, which will make no difference for the injured party other than a new lawsuit, or make them argue why the failure had not happened. The claimant can then claim that since it could never be caused by human error or a safety defect, it must have been a defect that the defendant is responsible for. As this is a civil lawsuit, in the situation where the defendant decides to deny all claims and not argue the claimant is likely to succeed, as silence on the matter speaks against the defendant considering the severity of the damage.

This is further supported by the role this and reimbursement outside of contract lawsuits present, since they will be used in situations without insurance.\footnote{We do not include considerations on insurance, as this is subject to contract and is very specific for each company that provides it in Danish law.}

The claimant will then have to prove the damage had occurred, which we assume they are able to, but they also have to prove the link between the defect and the injury, either a \emph{direct} or \emph{indirect} link. 

For the link to be direct, it has to be physically seen or decipherable from log files. For it to be indirect, it has to be derivable from the situation.

We have two cases that illustrate the duality of the indirect link. Lawsuits from the case law based approach are free to be used in the directive based approach, even if the legal sources for it are different.

In the case U.1939.16H, cattle owned by the claimant died after being fed black treacle. Out of a set amount of cattle, only those fed with the black treacle produced by the defendant died the following week. The claimant claimed, after having used an expert witness that showed that they were in good health before being poisoned, that they had not been overfed or otherwise damaged by the claimant, that the cattle which were not fed with it survived, and that poisoning from the black treacle could therefore be the only cause of death. The argument was built so because a vet at the time could not clinically prove it, which is why the link is indirect. Black treacle acts as supplement, and is not supposed to have any drawbacks. The defendant argued that the claimant had not proven this sufficiently, but did not provide additional reasoning for why this was so. The Danish Supreme Court found that there could be no other reason for the deaths, and sided with the claimant.

In the case U.2003.1706H, the claimant's roses that were fed with peat manufactured by the defendant experienced distorted growth. The parties agreed, that the distorted growth was caused by oxygen deprivation. An expert witness brought by the defendant made it clear that the contents of the peat were fit for use. The claimant used the same argumentation as from above, that is, that due to the circumstances, the peat causing it was the only plausible outcome. The Supreme Court found that just because the distorted growth stopped after a change in peat, it did not mean that the peat was the cause, nor that the peat could have been different than how it was described, and that the peat was not different than what was previously agreed and delivered between the parties. The court therefore sided with the defendant.

If we apply case law, we see that unless the claimant has access to log files that show adversarial failure, and/or design documentation that shows which defences and adversarial failures that were considered, the claimant should try to argue for an indirect link. 

They are likely to succeed, since the concept of the same product suddenly having a defect, the argument from the second case, does not apply to adversarial failures that are caused by inadequate defences. This is because defences are created to defend against threats, and requiring maintenance and updates is usually part of the service agreement on purchase.\footnote{However, if such an agreement is not part of the purchase of the surgical robot, the claimant will likely not be able to establish an indirect link.} The defendant has a case law based tool they can make use of, which is the test for whether the defect is ``systemic damage'',\footcite[477]{Obligation2015} which if true, shows that the product cannot be considered defective. The distinction between danger and defect is explained below, but has no effect on the test. The defendant would have to build their procedure around the test being fulfilled, which equates to two questions that must be answered with a yes. These are:

\begin{itemize}
    \item \emph{Are the dangers known?} The danger of manipulation attacks, subversion of robotic control and poisoning of the feedback are all known, but they are not necessarily known for each type of surgical robot. It has be known for the product. The claimant can argue, that they are not known and ask for documentation for this otherwise. The defendant can retort, that the underlying risk of any type of adversarial attack equates to public knowledge of the  dangers. But the defendant is unlikely to prove anything with such a general argument, since it is product specific, which we know from eg U.2015.572H\footnote{A groundbreaking case, where a claimant tried to sue the manufacturer of a big tobacco brand for the cancer that the excessive use of cigarettes had caused. They failed, because the damage caused to the claimant was considered systemic, because it was both known by everyone concerning the product, and unavoidable if you smoked it.} that 'known' refers to the product only, and vague statements are not accepted by the judges.

    \item \emph{Are the dangers unavoidable?} By unavoidable, it refers to whether the scientific and technical community deems it to be likely. 
    
    The defendant can claim that all adversarial failures are generally unavoidable because of new techniques and vulnerabilities, which is an argument of constant development. The claimant would retort that certain adversarial failures are more preventable than others. The exception would generally be subversion of robotic control, because the manufacturer cannot perfectly control the network that the surgical robot receives commands over.\footnote{This responsibility is incumbent on the hospital or any subcontractors that maintain networks and IT infrastructure.} They are however able to build suitable defences against the rest, even if doing so in CPS is difficult.

\end{itemize}

It is unlikely that both questions in this test can be answered with yes, since the public at large does not know that these adversarial failures can occur to surgical robots, but certain adversarial failures may be considered unavoidable. Most importantly, the judges have to be convinced of this, and even if both could be answered positively, that does not mean that the judges will decide to allow the test.

\subsubsection{Product liability lawsuit via case law}

Initially, it has to be mentioned that The Court of Justice of the European Union has concluded, that this approach can only be used where the product liability directive does not apply\footnote{See eg case C-183/00 \emph{González Sánchez} [2002] ECR 255.}where a Spanish set of product liability rules that put the patient/consumers in a more favorable position was ruled to violate the directive. \footcite[471]{Obligation2015} This approach can only be used on surgical robots that are not covered by the directive, see § 3 from the law of product liability. In practice this would limit it to completely custom made surgical robots, as well as those that cannot be moved without being destroyed\footnote{However, if EU law or the system ever changes, one can revert to relying on this case law based approach.} unlike the directive based approach.

While the test of systemic damage, and the use of case law from earlier, still apply to this approach, there are certain differences in how the defect is defined. Both the definition of defect from above can be used, as well as the old term ``danger''. \footcite[476]{Obligation2015} If the product can injure the user or third person, it is considered dangerous. But to be defective, it has to be ``unreasonably dangerous''.

This implies that some products are inherently dangerous to use, but it is the manner of danger outside of this that determine it. Candy as an analogy is not unreasonably dangerous, but excessive consumption may increase weight and damage health of individuals, but that does not make candy unreasonably dangerous. Any software system or device that is connected to the internet or local network poses a direct danger to the patient due to the known vulnerabilities and possibilities of abuse it contains. But for it to be unreasonable, it has to have a risk that occurs often or commonly, which so far seems not to be the case here.

The use of 'defect' from the directive is therefore appropriate, since unreasonably danger would likely not cover adversarial failures, and because the terms can be used interchangeably in both approaches.

Otherwise, the case would proceed as above.

\subsubsection{Lawsuit on product liability in contract based on case law}

The contractual approach is next, which is included for the sake of completeness. It was created from case law as well. If possible, this type of lawsuit would completely circumvent the rules laid out above, and instead merely focus on analogies to the Danish law of purchases, which would lead to cases where the evaluation of the sale of a proper product was met or not. This would apply to the patient as a third party, and allow a lawsuit. To make use of this, the claimant must first prove that there exists a contract between them and the operator or manufacturer. Clearly, the patient has not signed anything with either in written form, but has the patient done so orally? 

To assume the patient has accepted an oral contract with the hospital or the doctor, there has to be a so called “meeting of the minds” \footcite[21]{Gomard2012} in Danish law. Such a meeting must here include the acceptance of treatment being done in part or partially by a surgical robot, and the general risk of failure of the machine or anaesthetics. Whether they have to disclose the risk of adversarial failures is unlikely, since there has not been any such failure publicly recorded in Denmark. However, the claimant is not likely to prove that this exists, since Danish healthcare law does not work with contracts between these two parties in the context of private law. This means that the judge would dismiss the case on the basis of a lack of a contract.

\subsubsection{Comments on Evidencing}

As expressed above, proving that the defect exists in the product, and whether the link exists between the defect and the injury are complicated for the claimant. Initially, they do not posses the necessary design documentation or files, which show that the adversarial failure caused their injury. Like in other cases, they can show that nothing else could have caused the failure, such as via expert witnesses or even the operators that worked with the robot at the time, or system admins. This would force the defendant to provide some of documentation directly or otherwise leave the claimant unopposed.

Furthermore, the MDR has a special function in regards to proofs in these cases.\footnote{See section 4.4 of this paper.} In Art 10(14), paragraph 3, if the regulator determines damaged occurred, it can upon request transfer of all documentation that it has access to the patient/representatives, requiring there to be a public interest in disclosure to overrule any violation of data protection rights and without violating intellectual property rights. But since civil lawsuits in Danish law can be held behind closed doors, documentation that in an open case would violate those rights can be used if deemed possible by the regulator. 

The court decides on whether it is appropriate, if contested by the defendant. Public interest refers to what it literally reads as,\footnote{See Art 10(14), paragraph 3 of the MDR.} which means that it has to potentially affect more than the claimant, or have consequences otherwise, which adversarial failures on surgical robots likely warrant.

\subsection{Patient Compensation Association}

The act of complaint and reimbursement access in the health-sector\footnote{LBK nr. 995, 14/06/2018.} defines the structure and the requirements for Patient Compensation Association and cases processed by it.

The means to do so is called “Patient Compensation”, which is not a lawsuit, but instead an administrative process to receive compensation. This is facilitated by the Patient Compensation Association,\footnote{Staff includes doctors and lawyers.} which is financed and run by the Danish state and private parties,\footnote{Such as private hospitals.} and it solely considers and decides on cases in regards to patient injury.\footnote{See §§ 32 and 33 in the act.} It does so in the manner as any public authority would, via the principle of officiality\footnote{This principle is not codified, but seen in case law, literature and Danish Ombudsman's practice.} in Danish public law. It includes the collection of evidence by the authority if necessary, and perfect application of existing law and committing to the correct decision.\footnote{See the currently accepted definition by the Danish Parliamentarian Ombudsman \url{<https://www.ombudsmanden.dk/myndighedsguiden/generel-forvaltningsret/officialprincippet/>}, last accessed 14 December 2021.}

The patient applies for Patient Compensation on the association's website. This means that in this case, the only parties are the patient and the state, since the subject of such cases can only be the one who initiates it, and the state is by default the one who compensates. This means that this way to receive compensation completely excludes the manufacturer and costs them nothing.\footnote{If the manufacturers of surgical robots were also included as ``private parties'' and contributed to the Patient Compensation scheme, this would make Patient Compensation affect the manufacturer as well. We have been unable to confirm this.} 

The association does not publicly take recourse on manufacturers or private companies, but private hospitals and the like will need to use an insurance which may take recourse on them after compensating the patient.\footnote{See \url{<https://pebl.dk/da/om-os>}, last accessed 14 December 2021.}

The objective liability for this damage rest on the regions that run the hospitals, which in practice means the state.\footnote{See § 29 of the Act.}

The types of damages that are considered are laid out in § 20, part 1. Since it is not caused by the operator in our case, the damage sustained has to be caused by “errors or failures in technical apparatus”. Any type of failure, whether it was caused by neglect or other, are considered to be covered by this type of damage. The damage has to be caused “most likely” by the failure or neglect, see § 20 part 1. This is especially important, when the patient was considerably weakened, and the injury likely would not have caused any damage had they not been sick. The patient is advised to make sure, that the evidence necessary to prove that the surgical robot failed is seen by the authority. An adversarial failure in these kinds of systems is possible, and an injury caused by it always establishes a link between the failure and the injury. This style of 'link' is not unlike the one from the past section or the next, but still has its own case law, which shows how it is defined.\footnote{eg U.2011.1019H, where the death of a female patient, claimed to be caused by treatment with a bladder catheter at a hospital, was deemed unlikely because of the amount of diseases she already suffered from, which severely decreased her health. The lack of a link between the death and the specific treatment led to the dismissal by the Supreme Court.} But if the adversary is able to hide those details, the operator can still attest that the machine failed, so it would then go from a security to a safety case, of which there is well established practice that supports the patient, as well as the literal reading of § 20, part 1. This essentially means that the patient has two ways to prove § 20. After the application has been sent, the authority will collect information from the hospital where the injury took place, including documentation created by the operator of the surgical robot. 

The patient is free to provide further evidence, but since the authority has responsibility to make the right decision, they do not need to. If they want to, the patient is able to access\footnote{Equivalent to a Freedom of Information Request, or other national tools.} which documentation the authority base their decision on, and halt the process until they have provided further proof.

The decision can be appealed to the courts, or to the appeal board driven by the appropriate ministry.\footnote{Relevant ministerial organ, usually under the Minister of Health, but each government decides their own structure.}

The amount one is able to recover is comparable to a successful lawsuit, but the amount can be larger as there are no court or registration fees. Only compensation for quantifiable damage and pain and suffering is possible, as the Danish definition of tort is not directly applicable here.\footnote{See § 26 of the Law of Reimbursement Responsibility, LBK nr 1070 af 24/08/2018.} If the patient accepts the judgement, and does not act further, they will receive the compensation at the date of the decision.

\emph{In short}, as long as the patient sustained quantifiable damage, and the surgical robot has sustained a failure, even if it cannot be proven to have been caused by an adversarial event, they are able to receive compensation.

The caveat to this is the situation where adversarial failures become commonplace. This could lead to a Patient Compensation case not being possible\footnote{As it relies on the failure and unusual circumstances, and in reality, political willingness to compensate for the injures. This will hopefully never be the case.}, either because of a change in the act, or because the administrative practice would be interpreted differently.\footnote{Such as assuming that adversarial failures would not qualify.} 

\subsubsection{Comments on Evidencing}

Proving the injury and the adversarial failure occurred in the Patient Compensation environment is different, as seen with the influence of public law and the lack of court rules. The full burden of proof does not lie on the patient, but on the authority.

Since they merely need to decide on whether equipment encountered an error or failed, seeing the failure and damage occur would be sufficient. It is unknown whether they have individuals educated to read the logs that such adversarial failures will create, but they can require the hospital/manufacturer to explain this to them. Cooperation with the appropriate national regulatory agency\footnote{The Danish Medicines Agency in this case.} is possible as well, since the sharing of documentation between such parties is allowed, through the principle of officiality from earlier, as it includes gathering any information necessary and possible, and other specialised rules for sharing information between public authorities subject to the GDPR.\footnote{Regulation (EU) 2016/679 of The European Parliament and of The Council of 27 April 2016 on the protection of natural persons with regard to the processing of personal data and on the free
movement of such data, and repealing Directive 95/46/EC (General Data Protection Regulation) [2016] OJ L 119/1.}

\subsection{Reimbursement Outside of Contract}

The default for seeking compensation in Danish law, is reimbursement outside of law, where the four requirements, \emph{liability} (through acts of carelessness), \emph{quantifiable damage}, \emph{a link} between the responsible and the damage and that the link is \emph{adequate} have to be fulfilled.\footcite[23]{Eyben2013} Unlike the examples above, there is no objective responsibility, only \emph{culpa}\footnote{An equal term in English is carelessness.} which the manufacturer must have committed.

Traditionally the standard for what is not careless is what a \emph{bonus familias pater} would do. This is criticized in Danish law,\footcite[87]{Eyben2013} and the standard is gradually moving towards a focus on the breach of rules (both legal and otherwise) or the “normal right to act”. 

This is defined or at least elaborated on in case law,\footcite[85 - 88]{Eyben2013} but it has not specifically been done for manufacturers of CPS or surgical robots.

\subsubsection{Fulfillment of Criteria}

For the lawsuit to be successful, the patient that was injured by a surgical because of an adversarial failure has to make it likely for the court to find the criteria mentioned earlier fulfilled. 

First, the claimant has to show that the defendant acted carelessly in relation to the adversarial failure. There does not exist \emph{lex specialis} they can have broken here, which would normally be enough to show carelessness. The claimant will not initially have any documentation concerning what considerations were taken about the prevention of adversarial failures in the company. But they can bring forth the argument that the manufacturer failed to act to prevent the adversarial failure from occurring. 

This assumes this was made clear first, and if not, the claimant can say the same about the safety aspect that allowed the patient to be injured. The evaluation of carelessness is broad,\footcite[122]{Eyben2013} so the claimant could also bring out the danger of a surgical robot, which by its very nature warrants the utmost caution in both its design and later service. 

The defendant can either choose to bring out the documentation which shows that they did consider the one adversarial failure that caused the injury, but could also go for the approach where they do not reveal which one. Like in product liability, this is a disadvantage since silence or dismissal does not by itself prove anything in a civil lawsuit, unless the claims are unreasonable or out of proportion. Due to the severity of the injury and of the big risk it poses, as both the da Vinci and Magellan systems can cause internal hemorrhage, the barrier for carelessness is further lowered. The final nail in the coffin would be the support of ``responsibility based on profession''\footcite[12]{Eyben2013}
 that is seen in safety case law\footnote{To illustrate this tightening of the evaluation of culpa, we use U.2010.1350H. Here, the defendant had installed ventilation equipment at the address of the claimant, but a fire developed after the defendant had left the property. This had happened due to a known defect with smoke cartridges after the installation, and the defendant knew this from their own experience and otherwise. The judge concluded that because of this knowledge and because they are considered professionals in the business, they acted carelessly and were therefore responsible.}.

In a security context, we argue that if it is known that manipulation attacks, subversion of robotic control and poisoning of the feedback loop is possible, any professional (here manufacturers) must make all attempts to mitigate it, and must prove they have done so to not be held liable. The claimant is therefore likely to be able to fulfill this criteria.

A clear way to show carelessness is if rules have been violated or otherwise not complied with.\footcite[89 - 97]{Eyben2013} Even if the case is not about personal data, if the defendant had breached GDPR, the bar for proving what is mentioned is further lowered, but not as clear compared to if \emph{lex specialis} was breached.\footnote{There exists specialised legislation on reimbursement in specific situations in Danish law.} On the other hand, the defendant could bring out the argument that a third party (like the hospital) had not complied with security legislation such as the NIS directive, which would enable the defendant to prove, that if the breach of security by the hospital caused the adversarial failure or made it much more likely, they would not be liable. If this was the case, such a lawsuit would require the claimant, after having lost the initial lawsuit against the manufacturer, to sue the operator (hospital) of the surgical robot instead.

We assume that any injury sustained is quantifiable, which means that the second criterion is fulfilled. The claimant would need reports from hospital staff and could also get a second opinion on the injury. This will be contested by the defendant, who might bring an expert witness to scrutinize the documentation provided by the claimant, but this is rarely worth the costs.

The next issue becomes whether there is a link between the careless behaviour and the injury. 

The claimant has to prove that the careless behaviour directly or indirectly likely caused the damage. This is usually the most difficult part for the claimant to prove in a lawsuit. 

In reimbursement outside of contract, the claimant merely has to make it more than likely, that the carelessness caused the damage.

The defendant can attempt to prove that the injury would have occurred no matter what they had done, which brings us back to the different adversarial failures. If the failure is subversion of robotic control, there are situations where the manufacturer could not have implemented defences that could have prevented or mitigated it. In that situation, it would only be the network administrator and therefore the hospital that could have prevented it, which makes the lawsuit unlikely to succeed. But in regards to the other two, the claimant can argue that the chance of the adversarial failure occurring would have been lower had the defendant actively attempted to prevent it.

The defendant can retort that the injury would have occurred regardless of their behaviour. This is related to the idea of \emph{conditio sine qua non}, where the act of carelessness must have made it more than likely for the injury to occur. To do this, they would have the prove that the adversarial failure was impossible to prevent or mitigate, therefore making the link impossible to prove for the claimant. The burden of proof is also comparatively high, and would likely only apply to subversion of robotic control.

The defendant can highlight a tangential issue regarding the link, which is whether there are competing causes for the damage. They can argue that there always exists a risk for an adversary to cause an adversarial failure, regardless of their mistake, eg caused by the failure of software vulnerabilities in generic operating systems on their devices. The defendant would again need to prove this, and vague general statements about ``technical advancements'' and ``new techniques'' from hackers are too vague, and we argue that they do not constitute a competing cause.\footnote{This is because the vague statement is part of the background of the causes, as security as a branch accepts the risk of new developments, see \cite[313]{Eyben2013}.} Even if they can prove that an update of software they do control caused the failure, this does not equate to the judge supporting the argument, nor does it refute the compelling argument made by the claimant. In fact, this likely undermines their case, as they as the professionals must be able to handle updates to proprietary software that works in their manufactured equipment.\footnote{This area may change since CPS consists of many types of software constantly working together, and if adversaries abuse known vulnerabilities in proprietary OS or the like, to make a surgical robot suffer an adversarial failure, it may constitute a competing cause that could disprove the link between liability and damage. But the small line between disclaiming all responsibility for choices they themselves take (using software they have not developed in their own robot) and taking it is thin.}

A piece of safety case law can further show how difficult proving the link can be. 

In the case U.2011.354Ø, the ship of the defendant was captured by Somalian pirates. The claimants, the employees of the ship, claimed that the captain had not established increased surveillance of the dangerous waters, and had not taught the crew to use the alarms designed for these situation. The judge agreed, but found the defendant to not be liable, even if they had acted carelessly, since the capture would have occurred regardless.

If we apply this to our situation, it can be used by the defendant if they can prove, that the adversarial failures would have happened regardless of their careless behaviour. The analogy from pirates to adversaries is adequate, but requires that the defendant must reveal all details that could show that the adversarial attack was overwhelming enough to warrant them not being liable. This will in turn reveal which defences the defendant has deployed, which the claimant can use in other ways with expert witnesses, which makes this tactic risky.

Overall, if the adversarial failure was caused inside the surgical robot (and not via subversion of robotic control), the link between liability and the injury is likely to be proven in court, since a third party (the hospital) possesses local data that can likely show the failure or if the defendant reveals it as part of the process, or due to an indirect proof of it akin to the two examples of case law in section 5.1.1.

The last criterion is adequacy. Is it adequate that the link between the careless behaviour and the damage exists? This question is usually answered by case law, but adversarial attacks on surgical robots has not been considered by the Danish or Nordic courts yet. The claimant can argue that because the defendant is a professional party, with objective liability and a special role in both the product liability directive and the Medical Device Regulation, and because the harm is bodily, that the link is more adequate than not.\footcite[302]{Eyben2013} The defendant can then argue that they should not be responsible for any kind of attack directed at their produced machines, and that this should fall on the end user. Even if the defendant has contractually tried to abstain from any liability, this does remove the fact they are the only party capable of effectively preventing some of the adversarial failures, which would lead to the judge most likely dismissing the argument and contractual clause. The actions of third parties has been heavily discussed in the literature,\footcite[302]{Eyben2013} and it is clear that if a third party, such as the operator, caused the adversarial failure to occur by their actions, the defendant cannot be held liable. But situations where the action just made it more likely (and not guaranteed) does not mean that the defendant is off the hook. They must disprove this, easiest done with expert witnesses or compelling documentation that could show for eg show why the network security levels of the hospital were inadequate.

\emph{In short}, with a very high burden of proof, the patient can prove that the manufacturer acted carelessly in certain situations and therefore are liable, they can definitely show that an injury occurred, they can likely establish a link between their injury and the carelessness, albeit only some of the time, and that it is likely that it is adequate.

\subsubsection{Comments on Evidencing}

Like before, the most difficult parts of these lawsuits are proving the existence of the failure and whether the manufacturer is responsible.

We know that the judges are willing to reverse the burden of proof, especially when they feel like they have a very low chance of uncovering the “hidden proof”.\footcite[169]{Eyben2013} This refers to situations where the actors that cause the damage are wholly owned and used by the other party, which the claimaint would have no way of understanding or proving anything about.\footnote{This will always be the case regarding medical equipment in general, the patient should not have access or control over them.} This is also called the presumption of responsibility, and it leads to a situation where, if the defendant cannot prove that they did not act carelessly, they will be considered liable.\footcite[168]{Eyben2013} To reach this, the claimant must encourage the idea in the mind of the judge, that the necessary documentation they need will never come out of the defendant unless they make use of this principle.\footcite[169]{Eyben2013} But this is rarely used, and since we showed above that the case can most likely be decided without reversing the burden of proof, it is unlikely. But the claimant should use it in the situation where the case would fail on proving the link, since the reversal will require the defendant to prove they did not act carelessly, which is difficult under most circumstances.

The courts can also choose to tighten the evaluation of carelessness, or assume responsibility to be objective because of the circumstances, with a central case for this being U.1957.109H.\footnote{In it, a 14-year-old girl dropped out from an amusement park ride, and got injured. She did so because the back of the seat in the ride failed, and as she did not cause unusual strain to the seat, the supreme court concluded that the park was to cover her damages, since the seat was not strong enough for the task is was made for, and the park could not disprove this. While the judges at the time did not call it tightening of the evaluation of carelessness, it is later seen as such.}

This can be used by a claimant to argue that if the surgical robot and the infrastructure around it allow attacks that can cause bodily harm, they are not secure enough for their usage.\footnote{But the distinction between what the manufacturer is expected to be able to defend against, is still apparent here, see section 5.1.1 of this paper.}  If this argumentation is accepted, it allows the patient not to use resources to prove the liability.

\subsection{Adversarial Considerations}

Different adversaries come with their own issues.\footnote{See section 2.4.} Some may be so strong that their attacks amount impossible to defend against, and some give new opportunities for easing the burden of proof for the claimant/patient. We include some expanded comments on them here. Terrorist organizations, nation states, cybercriminals and organized criminal groups all require additional considerations. If the adversarial failures are induced by parties who are covered by criminal legislation, here terrorists/cybercriminals and organized criminal groups, the cases would be different in practice. The police and prosecutors would collect evidence, which would make the burden of proof for both patient and manufacturer considerably lighter, since reimbursement and product liability cases can make use of the evidence collected in criminal cases. Additionally, if organized criminals induced the failure(s), additional resources would be delegated to the investigation, and the potential punishments would be higher. Same goes for terrorists, since they are covered by anti-terror legislation. Both they and nation states as adversaries can cause \emph{force majeure}. This term in Danish law covers very unusual situations, where normal practice may not apply, which means the manufacturer is likely to not be kept liable. Stuxnet-like\footcite{Falliere2011} malware can be a concrete example of sophisticated malware, that may lead to situations where \emph{force majeure} could be used by the defendant. 

The mentioned malware is an example of an attack that led to all our mentioned adversarial failures, at once, and similar overwhelming attacks from nation states cannot reasonably be expected to be defended against. 

Acts of terror carry a similar connotation, and it would most likely lead to \emph{force majeure} situations, unless the attacks were simple and easily preventable.

A cybercriminal causing a failure is one thing, but a named terrorist organization causing it is completely different to a judge. The amount of resources available to prosecute organized criminal groups is higher as well, since they too have special legislation imposed upon them, which might also lead to a lighter burden of proof for the patient. 

These exceptional circumstance exceptions exist in other jurisdictions under different names, and may have similar adversarial specific implications, as do \emph{lex specialis} for specific adversaries. 

\subsubsection{Considerations on failures}

\emph{We defined} six different adversarial failures\footnote{See section 2.3 of this paper.} that uniquely fit surgical robots. 

We can divide them into 2 broad categories regarding where they hit, being internally and externally. Manipulation, reprogramming, misappropriation of trade secrets,\footnote{Misappropriation of trade secrets is an adversarial failure, because the adversary is able to compromise data and compromises the defences of the surgical robot - regardless if the action has no consequences in the short term.} poisoning the feedback loop and software vulnerabilities all exist inside the surgical robot. Subversion of robotic control is external, as this is an adversarial failure which occurs via the communication channel of the robot. 

This has legal consequences, as this becomes the hardest adversarial failure to defend against, and can therefore lead to litigation that will be ruled in favour of the manufacturer.\footnote{See section 5.1.1.} 

This is because the standard of what can reasonably be expected from a manufacturer, or from medical devices specifically,\footnote{See section 4.3.} is not full control over the communication channels, but merely the security of the surgical robot. It has to function as expected, and suffering adversarial or non-adversarial failures amounts to the opposite - within reason as subversion of robotic control causes these failures but must be mitigated by the hospital or anyone else controlling the network which the surgical robot uses.

\section{Additional Perspectives}

In this section, we discuss some key issues regarding the Patient Compensation Association and its use outside of Danish Law.

\subsection{Patient Compensation Association}

Instead of a lawsuit, the injured patient can choose to apply online for reimbursement for the damages caused to their body by an adversarial attack on a surgical robot.\footnote{See section 5.2.} The Patient Compensation Association is a public authority, and filing a case means that they will gather evidence and evaluate whether compensation is to be paid. It is according to the same standards as the lawsuits, but because the Association can both gather technical documentation from other authorities and testimonies from surgeons and log files from the surgical robot, the most important documentation is the proof that damage occurred. This is also because compensation should be paid regardless of why the surgical robot suffered a failure (adversarial or not), as all that has to occur is for the robot to fail.\footnote{See § 20 of the  Act.} This is by far the safest and quickest way to receive compensation, and because of the argument above, it disregards any security issues and instead makes it about safety - which severely reduces any theoretical burden of proof that the patient may have had in regards adversarial attacks in the lawsuits. 

\subsubsection{International Perspectives}

Surgical robots are always at the risk of suffering an adversarial failure. Your surgeon will not make mistakes because they were hacked, but a surgical robot operating on you will. For the best interests of the patient, and within or outside the opportunities of the MDR,\footnote{See Art 10(16).} taking the best practice from the Danish legal system would be advantageous such as on a international scale. A sort of Patient Compensation system implementation everywhere where surgical robots are widely used.\footnote{Increased and clearer rights regarding the special situation that robotics in general put us in are supported academically, see eg \cite[30 - 33]{Leenes2017}.} One can reduce costs for the individual, reduce costs for the manufacturer and streamline the legal process and ease the means of redress for the patient. The clear disadvantages is that the reimbursement amount could theoretically be lower than that gained from lawsuits, but this can be solved by legislatively forcing the proposed system (like the Danish one) to use the same legal principles to calculate costs. Another interesting feature is the blatant disregard of technology - what matters is that the machine failed, adversary or not. This makes it a safe technology neutral approach to an otherwise very technology specific problem, and acknowledges the shift from safety to security problems.\footnote{Note that many jurisdictions' use of tort law/reimbursement law already disregard technology, but that this may not be to the advantage of claimants or patients.} It also removes most issues with burden of proof that the patient the would have had, had they taken their cases to court. 

\section{Future Work}

If adversarial failures in the future become commonplace, it might lead to it not being considered a defect or a failure in the Patient Compensation system or similar. Considerations on this topic outside or inside of the CPS or robot sphere or as interdisciplinary work is therefore highly needed. 

Analysis on when an adversarial attack constitutes a defect in different bodies of EU member states' law is necessary as well, since it is not clear whether the product liability directive addresses this. A general discussion of adversarial failures and defects on an international level is needed too.

Another area that needs further work is the MDR. Even if it itself declares that it has not changed significantly since its directive form in preamble 4, but rather has been reinforced, this does not mean that it will not be applied differently in practice. 
 
Research into what constitutes accessories to medical devices is necessary too, as well as which responsibilities the manufactures of the accessories have, regardless of future guidelines or case law. Increased complexity of CPS as medical devices like surgical robots, means and increasingly more complex field of liability.

\section{Conclusion}

In this paper, we initially created six distinct adversarial failures for surgical robots, to conceptualise and define which types of failures they may suffer from an adversarial attack, alongside five types for non-adversarial failures. The six adversarial failures are: Manipulation attacks, subversion of robotic control, reprogramming of the robot, misappropriation of trade secrets, poisoning of the feedback loop and software vulnerabilities.

We then showed how the MDR assigns liabilities and rights to both manufacturers of surgical robots and regulators. Following this, we showed how a manufacturer of a surgical robot could be kept liable for the damage caused to a patient by an adversarial attack in Danish law, through two types of lawsuits and a public law controlled non-court process. We showed that the European Medical Device Regulation that governs surgical robots does not on its face consider security aspects. There are considerations about the health and safety of patients, but not specifically about the risk that adversarial attacks pose.\footnote{However, this is understandable due to the technology neutral nature of the regulation.} Only the guidance that comes with the regulation, as well as an expanded interpretation on its rules of the risk and quality management systems come close to outright requiring a focus on security. 

As for other EU legislation, there is a possibility that the EU can take subsidiary measures to address out what the regulation misses.\footnote{See preamble 101.} 

We interpret several obligations for manufacturers as including security, and since they explicitly require elimination of risks and security levels, and proper functioning of the medical devices\footnote{Which means that adversarial attacks must be mitigated so as to make the devices work as intended at all times.} they have a higher chance of working in a cumulative manner, and ensure security despite its more general wording. 

It furthermore burdens the manufacturer (because of the increased obligations), which if breached when it comes to security and defences in the surgical robot, can be used effectively to support careless behaviour in civil litigation.

Furthermore, there is the possibility of future legislation down the line, as well as a the current rights of regulators to inspect, withdraw and generally keep a close eye upon the surgical robots if they are willing to do so.\footnote{See eg Art 10(14) in the MDR.} 

The regulators' use of rights has yet to be seen regarding adversarial attacks on surgical robots, and while useful, there is no guarantee that the regulators have the staff or finances for it.\footnote{But given how the new EU AI draft regulation can require specialised staff and expertise in its authorities, see Art 59 of the draft, this seems very likely.} Regarding the application of Danish Law to the situation of a surgical robot suffering an adversarial attack, we find that the issue of proving anything in court can be a major obstacle for lawsuits. Design documentation, log files and other documentation that the manufacturer has access to, is not initially able to the patient. 

But because civil lawsuits rely on free argumentation from both parties, the patient can indirectly force such proof out,\footnote{Through case law, see section 5.1.1 of this paper.} or via the MDR.\footnote{See Art 10(14).}

We find a lawsuit based on product liability possible, if it is based on the EU directive or case law based approach in Danish law,\footnote{To use the case law approach for litigation, the Product Liability Directive must not be applicable, see section 5.1.2 of this paper.} but not if it is based on contract.\footnote{This approach requires there to be a contract between the patient and the manufacturer of the surgical robot, which there is not, but future case law could reveal a new way to interpret this approach.} 

The two useful types of lawsuit require that the surgical robot is put into circulation, so completely custom made versions are exempted. 

Especially the use of \emph{res ipse loquitur}, which is showing that nothing else but an adversarial failure could have caused it is likely a very efficient approach in court. 

This, supported by case law, shows the link between the defect and the injury, but the defendant has one last defense they can ask for which is the case law based test of ``systemic damage''. 

If the danger adversarial attacks pose are known and unavoidable, the defect can be disproved and the patient will most likely lose. 

But both questions have to be answered negatively in most situations, because the risk of attacks are not known by the public for the product, and only subversion of robotic control as an attack can be considered unavoidable. 

We show that the patient is able to sue for damages via reimbursement outside of contract, but proving the link between the attack and the injury is difficult. Indeed, if the attack was subversion of robotic control, which involves factors outside of the surgical robot itself, the link is likely impossible to prove. And the patient can further attempt to argue that the needed knowledge is kept so closely to the other party, that it would be better shown if the judge reverses the burden of proof, which would bypass all needs to prove any criteria necessary to use this approach and instead force the manufacturer to prove that the surgical robot is designed appropriately, which is a direct reversal of the burden of proof.

Finally, the Patient Compensation system is likely a model way to cover damages. Instead of suing the manufacturer of the robot that was attacked, the patient can choose to submit a free application to the Danish Patient Compensation Association, and get their damages fully covered.\footnote{If they were deemed to have been injured.} This is only between the state and the patient, which means it is cheaper and easier for the manufacturer of a surgical robot. It is sure to succeed because the rules surrounding this dictate, that if the machine fails, no matter the cause, the patient is entitled to have all their damages fairly covered.

We can illustrate which remedies are likely possible and which are likely not:

\begin{center}
    \begin{tabular}{ | l | l | l | p{3cm} |}
    \hline
    Law & Product Liability & Patient Compensation & Reimbursement in court \\ \hline
    Unlikely & 0 & 0 & 0 \\ \hline
    Likely & X & X & X \\ \hline
    Highly likely & 0 & X & 0 \\\hline
    \end{tabular}
\end{center}

In terms of lessons for other systems, we argue that the model seen in the Patient Compensation system in Denmark is a highly appropriate mechanism with the issues adversarial attacks on surgical robots present, as it is technology neutral, equates safety issues with security and provides an efficient way for a patient to receive compensation from the damage caused by adversarial attacks on surgical robots which harmed them. 

\newpage

\printbibliography

\end{document}